\documentclass[journal]{IEEEtran}

\usepackage{graphicx}
\usepackage{amsmath}
\usepackage{amssymb}
\usepackage{color}
\usepackage{verbatim}
\usepackage{url}
\usepackage{bm}
\usepackage{multirow}
\usepackage{booktabs}
\usepackage{rotating}
\usepackage{lipsum}
\usepackage{cite}
\usepackage{ulem}
\usepackage[european]{circuitikz}
\usepackage{float}
\usepackage{subfig}
\ifCLASSINFOpdf
\else
\fi
\begin{document}
	%
	\title{Modeling of Organic Metal-Insulator-Semiconductor Capacitor}
	%
	%
	%
	
	\author{Prashanth~Kumar~Manda,~
		Logesh~Karunakaran,~
		Sandeep~Thirumala,~
		Anjan~Chakravorty, ~\IEEEmembership{Member,~IEEE,}
		and~Soumya~Dutta, ~\IEEEmembership{Member,~IEEE,}
		\thanks{Prashanth Kumar Manda, Logesh Karunakaran, Anjan Chakravorty, and Soumya Dutta are with the Department of Electrical Engineering, Indian Institute of Technology Madras, Chennai 600036, India.}
        \thanks{Sandeep Thirumala was with the Department of Physics, Indian Institute of Technology Madras, Chennai 600036, India (e-mail: s.dutta@ee.iitm.ac.in).}
        \thanks{The authors would like to acknowledge Department of Science \& Technology (DST, Govt. of India) and IIT Madras for financial support.}}
	\maketitle
	
	\begin{abstract}
		In this paper, we demonstrate the principle of operation of a metal-insulator-semiconductor (MIS) capacitor based on undoped organic semiconductor. In spite of low charge concentration within the semiconductor, this device exhibits a capacitance variation with respect to applied gate voltage $V_g$, resembling the capacitance-voltage $C$-$V$ characteristics of a traditional doped semiconductor based MIS capacitor. A physics based model is developed to derive charge concentration, surface potential $\psi_s$ and the capacitance of organic MIS capacitor. The model is validated with TCAD simulation results and is further verified with experimental results obtained from fabricated organic MIS capacitor consisting of poly(4-vinylphenol) and poly(3-hexylthiophene-2,5-diyl) as insulator and semiconductor, respectively. 
	\end{abstract}
	

	%
	\IEEEpeerreviewmaketitle

	\section{Introduction}
	%
	%
	%
	%
		\IEEEPARstart{M}{etal} -insulator-semiconductor capacitors are considered to be a model device structure as a preamble to study the field effect transistors. Substantial progress of organic field-effect transistor, which consists of undoped organic semiconductor as an active material, strongly invokes the analysis of MIS capacitors. Till now there have been few reports on the capacitance-voltage ($C$-$V$) characteristics of organic semiconductor based MIS capacitors, which have been explained based on doping concentration \cite{DitDetrmine1980,ModelTED,TaylorphtoAPL} and empirical models \cite{MStauMeijerAPL2001,TaylorInterfaceJAP2008,Taylorinterface2005,MeijerMSvalidityAPL2001} under the framework of Mott-Schottky analysis. However, Mott-Schottky analysis, which has been widely accepted to explain $C$-$V$ characteristics of doped semiconductor based MIS capacitor, cannot be employed in undoped semiconductor based MIS capacitor as it leads to erroneous results, reported by \cite{Nigam1}. Thus there has been no report on physics based model describing $C$-$V$ characteristics, which is indispensable to understand the principle of operation of organic semiconductor based MIS capacitor.
	
	In this work, we develop a physics based analytical model to explain $C$-$V$ characteristics of MIS capacitor, consisting of organic semiconductor, which is typically intrinsic in nature. In the context of the present model, the mathematical expression of charge profile, surface potential and the capacitance are evaluated enabling an equivalent circuit model that can explain the observed $C$-$V$ characteristics of organic MIS capacitor along with its variation with semiconductor thickness. The model is validated with TCAD simulation and is further verified with experimental results that are obtained from the organic MIS capacitor, fabricated in our laboratory.
    
    The schematic of an organic MIS capacitor and its typical $C$-$V$ characteristics are shown in Figs. 1(a) and (b), respectively. Based on the applied gate voltage ($V_g$) and the flat band voltage ($V_{fb}$), $C$-$V$ characteristics can be divided into three regions: 1) $V_g << V_{FB}$ (strong accumulation region), 2) $V_g < V_{FB}$ (moderate accumulation region) and 3) $V_g > V_{FB}$ (weak accumulation region). In strong accumulation regime, the capacitance approaches to insulator capacitance ($C_i$). On the other hand, in weak accumulation regime the capacitance reduces to $C_{min}$, which is a series combination of $C_i$ and geometrical capacitance of the semiconductor ($C_{sc}$), as given by \eqref{Eq1C_cicg}  \cite{Nigam1}.     
    \begin{equation}
	C_{min} = \frac{C_i  C_{sc}}{C_i +  C_{sc}},
	\label{Eq1C_cicg}
	\end{equation}
	\begin{equation*}
	C_i = \frac{\varepsilon_i \varepsilon_0}{t_i},\quad  C_{sc} = \frac{\varepsilon_s \varepsilon_0}{t_s},
	\label{Eq1C_ci}
	\end{equation*} where $\varepsilon_0$ is the permittivity of free space, $\varepsilon_i$, $\varepsilon_s$ correspond to relative dielectric constants of insulator and $t_i$, $t_s$ are thicknesses of insulator and semiconductor, respectively.
    
However, there is no model to explain the capacitance behavior in the moderate accumulation regime in consistent with the regions 1 and 3. This work focuses on developing physics based model to describe the capacitance behavior covering the entire voltage range (regions 1-3) under consideration.	
	\begin{figure}[t] 	
		\centering
		\subfloat{\includegraphics[scale =0.21, trim = 0mm 0mm 0mm 0mm ]{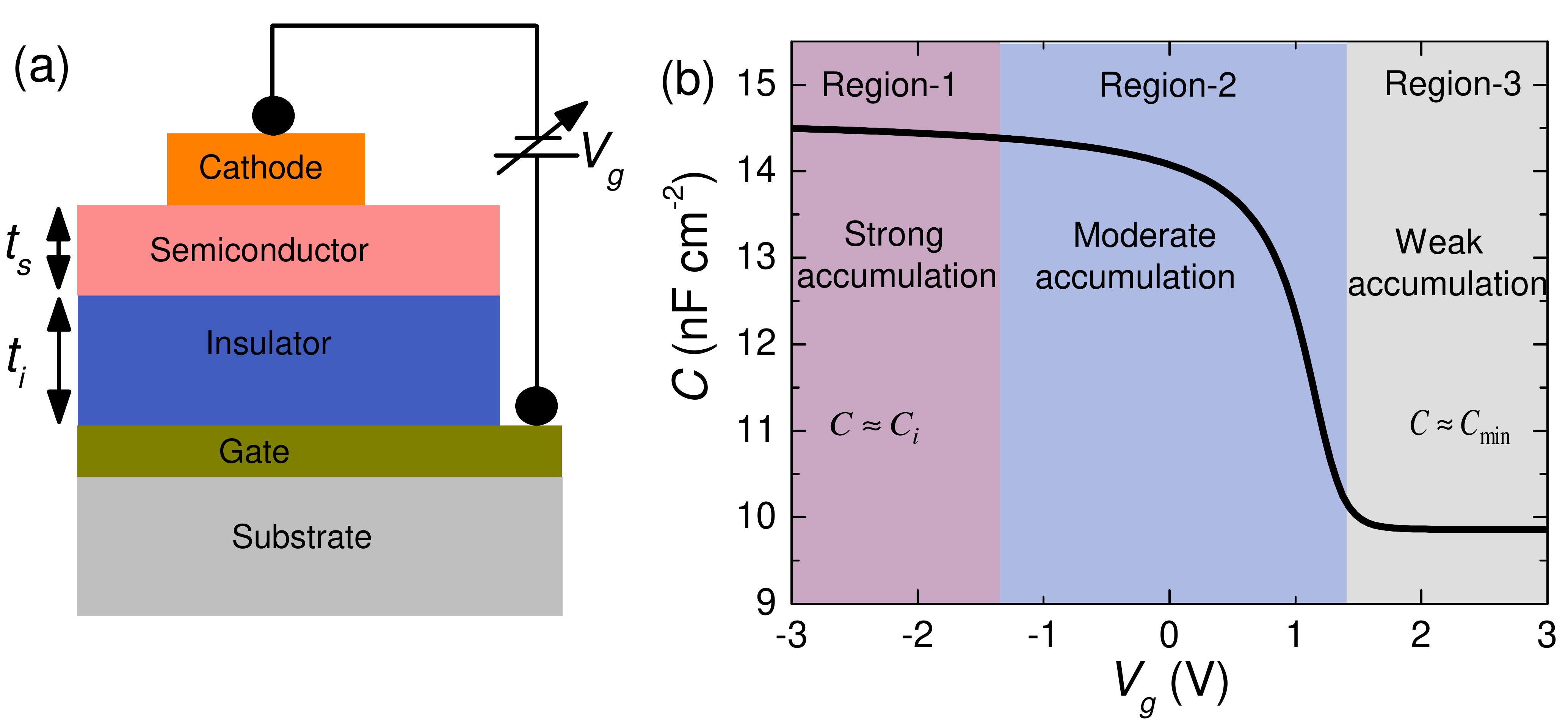}}
		\caption{(a) Device structure, (b) $C$-$V$ characteristics of a MIS capacitor.} 
		\label{Fig1OMIS_structure}
	\end{figure}	
	 
	\begin{figure}[t] 	
		\centering
		\includegraphics[scale =0.4,trim = 25mm 60mm 20mm 43mm ]{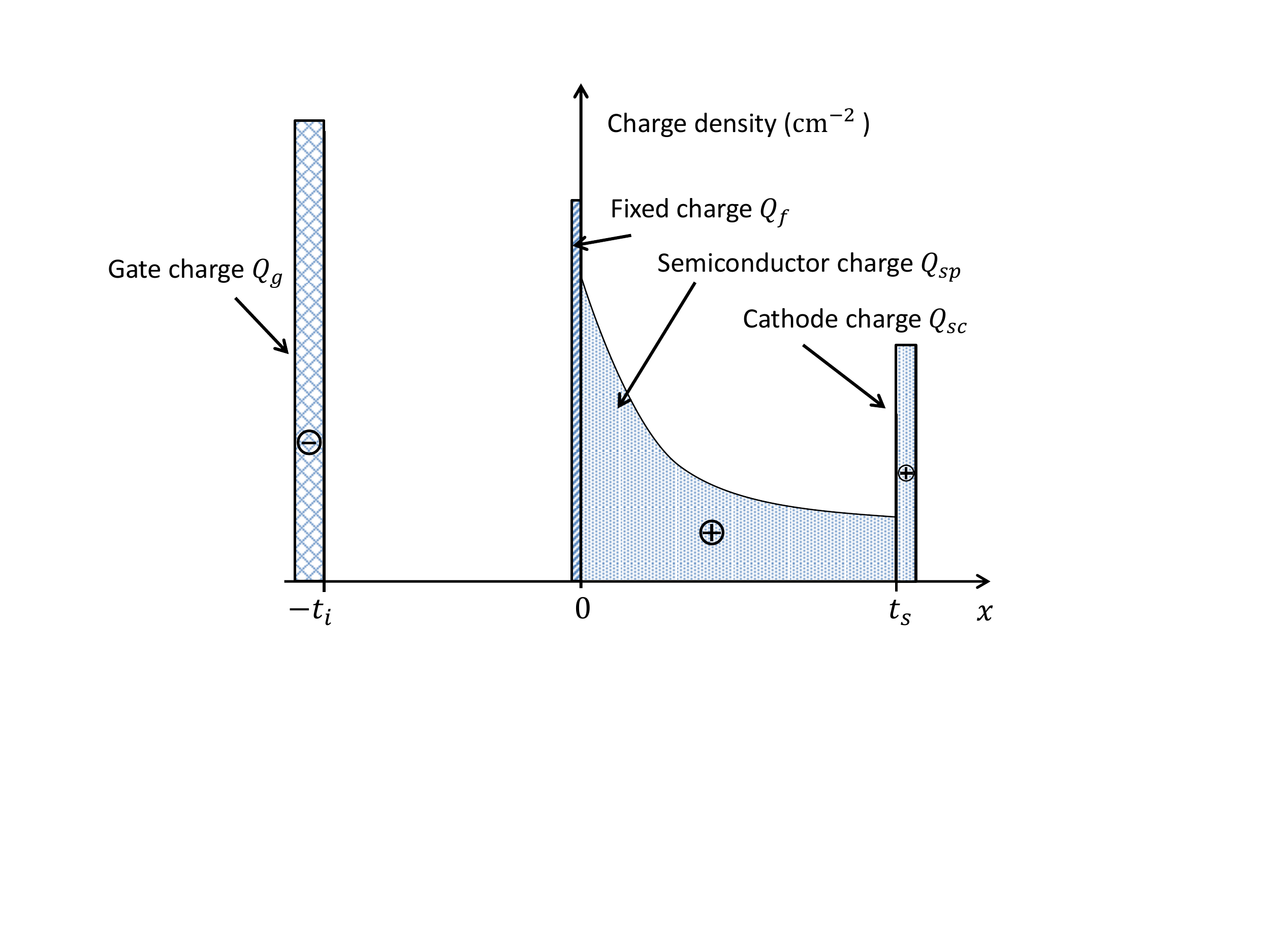}
		\caption{Schematic of the different charge densities.} 
		\label{Figchargeschematic}
	\end{figure}
	\section{Model development} \label{SecCapclc}
	In this study, we consider an intrinsic organic semiconductor, which forms a Schottky type contact with the cathode metal \cite{Manda,Nigam1,Nigam2roleofinject}, having a large barrier height $\phi_e$ (= 1.72 eV) for electrons compared to that for holes $\phi_h$ (= 0.28 eV). Upon formation of the contact, the charge carriers (electrons and holes) are injected into the semiconductor via thermionic emission process maintaining electron ($n_d$) and hole ($p_d$) carrier concentration at the semiconductor-cathode interface depending on the barrier height \cite{Manda,Nigam1,Nigam2roleofinject}. Since $\phi_h$ $\ll$ $\phi_e$, the injected hole concentration is more than the electron concentration (i.e., $p_d$ $\gg$ $n_d$) and the contribution of electrons can be neglected. The injected holes undergo diffusion due to concentration gradient and drift due to the electric field present inside the semiconductor simultaneously, establishing a steady state charge concentration profile under equilibrium. 
	
	Upon applying negative $V_g$, positive charges get induced on the cathode metal ($Q_{sc}$) as shown in Fig. \ref{Figchargeschematic}. Subsequently the injected holes are attracted towards the insulator-semiconductor surface resulting hole accumulation within the semiconductor ($Q_{sp}$ =$\int_0^{t_s}\! qp(x) \,\mathrm{d}x$, where q is the charge of electron). Different charge densities present within the device are illustrated schematically in Fig. \ref{Figchargeschematic}, where $Q_g$ and $Q_f$ represent the charge density at gate metal and the fixed charge density at the insulator-semiconductor interface, respectively. The charge density at the cathode metal $Q_{sc}$ can be expressed in terms of the potential drop across the semiconductor ($\psi_s$) and the capacitance $C_{sc}$ as 
	\begin{equation}
	Q_{sc}=-\psi_sC_{sc}. 
	\label{EqQc}
	\end{equation}
Using Kirchhoff's voltage rule, applied $V_g$ can be related to $\psi_i$ and $\psi_s$ as
	\begin{equation}
	V_g  = \psi_i +\psi_s. \label{EqVgVs}
	\end{equation}
	where $\psi_i$ can be replaced by  $-(Q_{sp} + Q_{sc} )/C_i$, in the presence of $Q_f$ and $\phi_{gc}$ (work-function different between gate and cathode) \eqref{EqVgVs} can be modified to
    \begin{equation}
	V_g -V_{fb} = \psi_s - \frac{Q_{sp} + Q_{sc} }{C_i}, \label{EqVgVsQs}
	\end{equation} where $V_{fb}$ = $\phi_{gc} - Q_f/C_i$.
 
 Differentiating \eqref{EqVgVsQs} with respect to $Q_g$ one obtains
	\begin{equation}
	\frac{1}{C} = \frac{1}{C_{i}} +\frac{1}{C_{sp}+C_{sc}}.
	\label{EqTotalC}
	\end{equation}
where \begin{equation}
C=\partial Q_g/\partial V_g,
\end{equation} 
 \begin{equation}
 C_{sp}=-\partial Q_{sp}/\partial \psi_s,
 \end{equation} \begin{equation}
\partial Q_g = -\partial (Q_{sp} + Q_{sc}).
\end{equation}

\begin{figure}[t]
		\begin{center}
			\begin{circuitikz}[scale=0.8]
				\draw (0,0)	to[C=$C_i$,o-] (2,0);
				\draw (2,0)	to[short] (2,1) 
				to[short] (2,-1);
				\draw (2,1)	to[C=$C_{sc}$] (5,1);
				\draw (5,-1)to[C=$C_{sp}$] (2,-1);
				\draw (5,-1)to[short] (5,1);
				\draw (5,0)	to[short,-o] (6,0);
			\end{circuitikz}
			\caption{Equivalent circuit diagram according to \eqref{EqTotalC}.}
			\label{CktEq}
		\end{center}
	\end{figure}
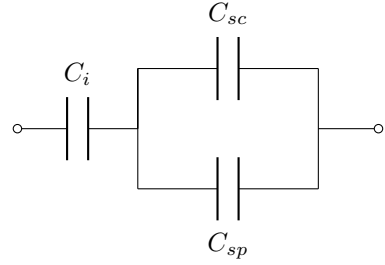

Equation \eqref{EqTotalC} can be manifested by an equivalent circuit, representing organic MIS capacitor, as shown in Fig. \ref{CktEq}. It is evident from \eqref{EqTotalC} that the capacitance variation with $V_g$ (especially in region 2 of Fig. \ref{Fig1OMIS_structure}(b)) is due to $C_{sp}$, arising from the variation of injected mobile charge carriers with respect to $V_g$. Thus a complete model for capacitance of organic MIS capacitor can be achieved by modeling $Q_{sp}$ (so $p(x)$) and $\psi_s$.	
    
    A coupled equation for the electric field ($E(x)$) within the semiconductor $\left(0\leq x\leq {t_s}\right)$ can be obtained by combining Poisson's equation $\left(\partial E(x)/\partial x = q ~p(x) / \varepsilon_s \varepsilon_0 \right)$ along with the hole transport equation under equilibrium as 
	\begin{equation}
	\\ \dfrac{E(x)\partial E(x) }{\partial x} - V_t \dfrac{\partial^2 E(x) }{\partial x^2}=0 , \label{CoupledEx}
	\end{equation}
	where $V_t$ is the thermal voltage. Equation \eqref{CoupledEx} is integrated once to get 
	\begin{equation}
	\\ \left(\dfrac{1}{2 V_t}\right)^2 E(x)^2 - \dfrac{1}{2 V_t} \dfrac{\partial E(x) }{\partial x}=k^2, \label{IntCoupledEx}
	\end{equation}
	with $k$ as a constant of integration. For a semiconductor with finite thickness, a general solution for  \eqref{IntCoupledEx} was given by \cite{Skinner} as 
	\begin{equation}
	E(x) = - 2 V_t k \coth\left(k\left[x- 2 V_t \xi\right]\right).\label{Egensol}
	\end{equation}
	where $\xi$ is a constant to be evaluated. Subsequently, $p(x)$ can be written as 
	\begin{equation}
	p(x) = \frac{2 \varepsilon_s \varepsilon_0 V_t}{q} \left[\dfrac{k}{\sinh\left(k \left[x-2 V_t \xi \right]\right)}\right]^2.\label{pgensol}
	\end{equation}
	Using the boundary condition at the semiconductor-metal interface ($p(t_s)=p_d$), \eqref{pgensol} further takes the form of
	\begin{equation}
	p(x) = p_d \left[\dfrac{k x_d}{\sinh\left(k \left[x-t_s\pm x_1\right]\right)}\right]^2,\label{pgensolx1}
	\end{equation}
	where $x_1 = \dfrac{\arg\sinh\left(k x_d\right)}{k}$ and $x_d =\sqrt{\dfrac{2 \varepsilon_s \varepsilon_0 V_t}{q p_d}}$.
	
    By integrating $E(x)$ and $p(x)$ within the semiconductor, $\psi_s$ and $Q_{sp}$ can be obtained respectively, as given below 
	\begin{equation}
	\begin{aligned}
	\psi_s 	= 2 V_t \log\left(\dfrac{ \sinh\left(k\left [\pm x_1-t_s\right]\right)}{\sinh\left(k\left [\pm x_1\right]\right)}\right).
	\label{Eqpsi_s}
	\end{aligned}
	\end{equation}
	 
	\begin{equation}
	Q_{sp}(\psi_s) = Q_0 \left(\sqrt[]{\exp\left(\frac{-\psi_s}{V_t}\right) + (k x_d)^2} - \sqrt[]{1 + (k x_d)^2}\right).
	\label{EqQ_s_anlt} 
	\end{equation} where 
    \begin{equation*}
    Q_0 = \sqrt[]{2 q \varepsilon_r \varepsilon_0 V_t p_d}
    \end{equation*}
	It is to be noted that the unknown parameter $k$ needs to be evaluated to calculate $Q_{sp}$ and $\psi_s$. $k$ can be calculated by solving \eqref{EqVgVsQs} with the help of \eqref{Eqpsi_s} and \eqref{EqQ_s_anlt}. The developed model is valid for $x_d>t_s$.
    
    For $x_d<t_s$ (is realized by reducing the barrier height for holes )and in strong accumulation, the charge concentration at $x=d$ becomes significant and comparable to the charge concentration within the device as shown in Fig. \ref{Fign_E_Q_Vs_HSC}(a). Moreover the charge concentration is sufficient to screen the gate field as shown in Fig. \ref{Fign_E_Q_Vs_HSC}(b) and gate looses the control over the cathode charge. Hence in this case the cathode charge is invariant with the gate voltage.       
	
	\begin{figure}[t] 	
		\centering
		\includegraphics[scale =0.2, trim = 15mm 0mm 0mm 0mm  ]{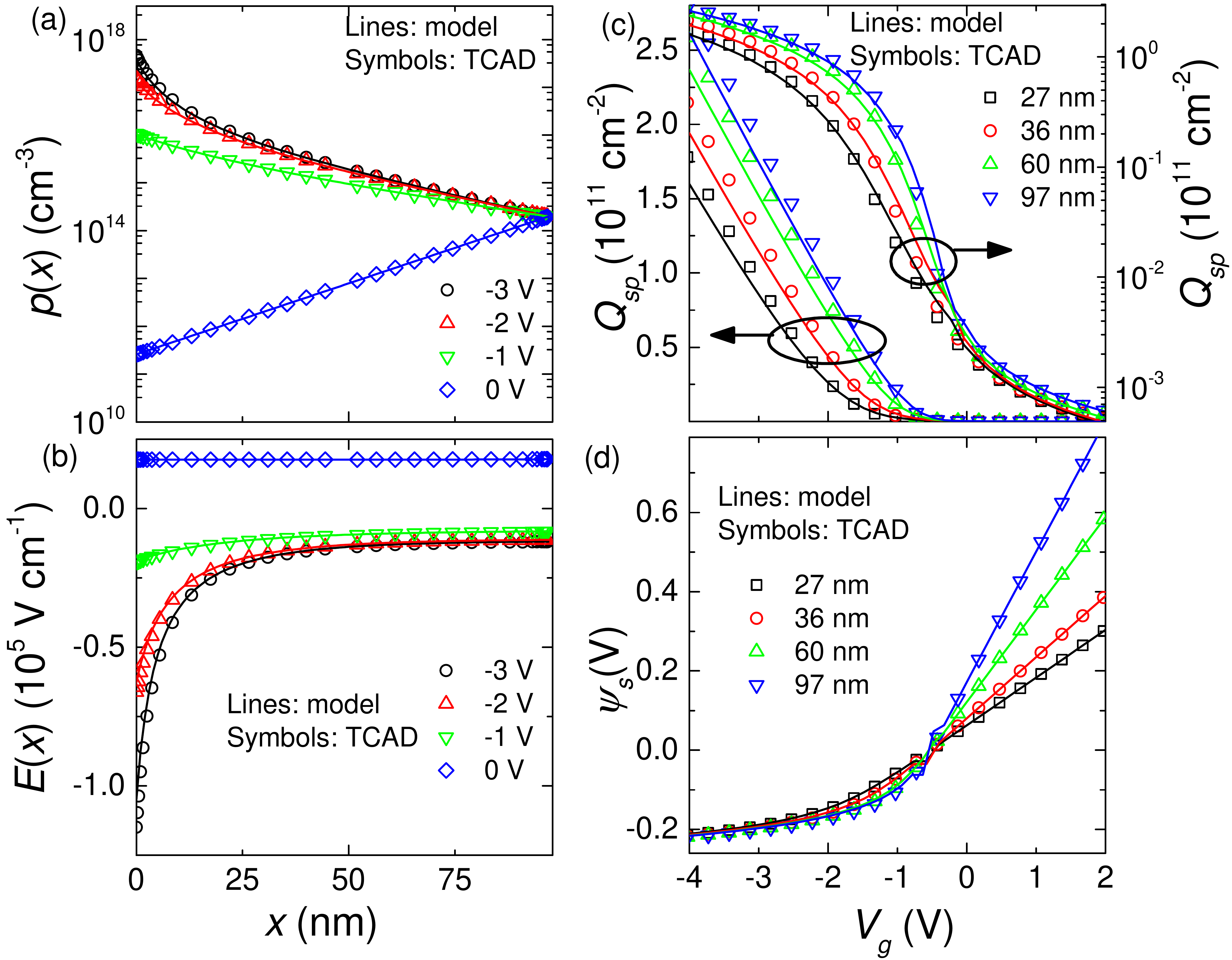} \\
		\caption{(a) $p(x)$ and (b) $E(x)$ for different $V_g$ with $t_s$ = 97 nm, (c) $Q_{sp}$-$V$ characteristics in linear (left axis) and logarithmic (right axis) scales and (d) $\psi_s$ variation with $V_g$ for different $t_s$ for $x_d>t_s$ case. The symbols are TCAD simulation and solid lines are model. The parameters used for the simulation are $t_i$ = 287 nm, $\varepsilon_i$ = 4.81, $\varepsilon_s$ = 3.3, $\phi_h$ = 0.28 eV, $N_C$ = $N_V$ = $10^{19}$ cm$^{-3}$ and $V_{fb}$ = -0.52 V.} 
		\label{Fign_E_Q_Vs}
	\end{figure}
		 \begin{figure}[h] 	
		\centering
		\includegraphics[scale =0.2, trim = 15mm 0mm 0mm 0mm  ]{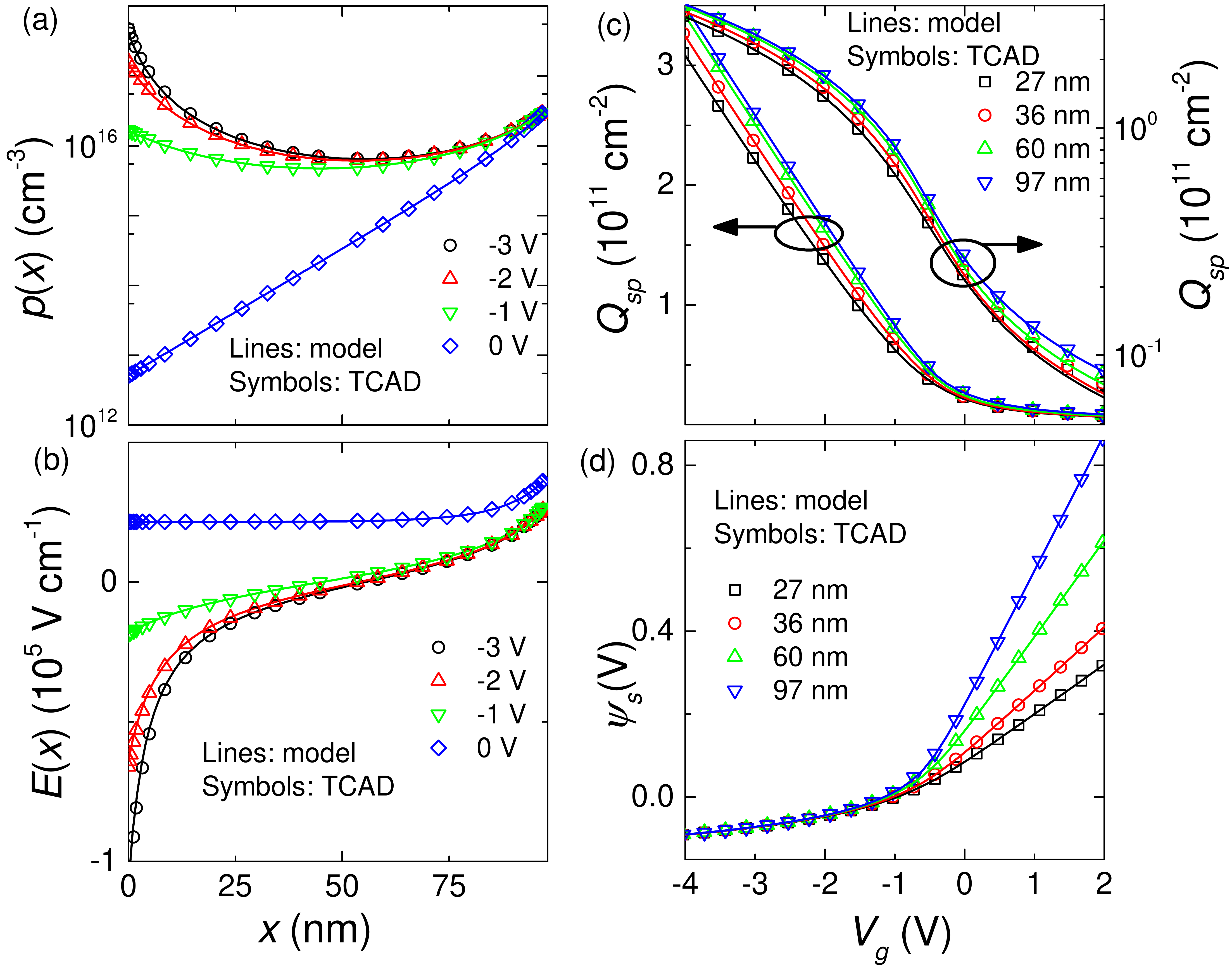} \\
		\caption{(a) $p(x)$ and (b) $E(x)$ for different $V_g$ with $t_s$ = 97 nm, (c) $Q_{sp}$-$V$ characteristics in linear (left axis) and logarithmic (right axis) scales and (d) $\psi_s$ variation with $V_g$ for different $t_s$ for $x_d<t_s$ case with $\phi_h$ = 0.15 eV, and $V_{fb}$ = -0.52 V. The symbols are TCAD simulation and solid lines are model.} 
		\label{Fign_E_Q_Vs_HSC}
	\end{figure}
\section{Model Validation and Discussion}
	The model, developed in the preceding section, is validated using Sentaurus TCAD simulation \cite{Sentaurus}.  The variation of hole concentration profile with respect to $V_g$, calculated from the model (solid lines), is in good agreement with the TCAD simulation results (symbols), as shown in Fig. \ref{Fign_E_Q_Vs}(a). It is evident that the hole concentration at $x$ = $t_s$ is bias independent as it is solely decided by thermionic emission. On the contrary, the hole concentration at the insulator-semiconductor interface $\left(x=0\right)$ varies exponentially with $\psi_s$ as 
    \begin{equation}
    p(0) = p_d \exp\left(\frac{-\psi_s}{ V_t}\right).
		\end{equation}
Similarly, the variation of electric field profiles with $V_g$ across the semiconductor are plotted in Fig. \ref{Fign_E_Q_Vs}(b) showing a good match between the model (lines) and TCAD simulation results (symbols). The constant electric field profile throughout the semiconductor for $V_g>V_{fb}$ supports the fact that at higher voltages the semiconductor resembles an insulator.   
    The variations of $Q_{sp}$ and $\psi_s$  with respect to $V_g$ for different semiconductor thickness $(t_s)$ are shown in Figs. \ref{Fign_E_Q_Vs}(c) and (d), respectively. Upon decreasing $V_g$, more holes are attracted towards the interface, which results in an increase in $Q_{sp}$ and a decrease in $\psi_s$, as expected from \eqref{EqQ_s_anlt}. In contrast,  with the increase in $V_g$, holes are repelled from the insulator-semiconductor interface making $Q_{sp}$ much lower than $Q_{sc}$. This turns the semiconductor to behave like an insulator (Fig. \ref{Fign_E_Q_Vs}(b)). Consequently $\psi_s$ increases linearly with $V_g$ (Fig. \ref{Fign_E_Q_Vs}(d)), which can be understood from \eqref{EqVgVsQs} by neglecting $Q_{sp}$ term and replacing $Q_{sc}$ in terms of $\psi_s$ using \eqref{EqQc}. The dependences of $Q_{sp}$ and $\psi_s$ on semiconductor thickness, as expected from \eqref{EqQ_s_anlt} and \eqref{Eqpsi_s}, are clearly observed in Figs. \ref{Fign_E_Q_Vs}(c) and (d) respectively. It is to be noted that all $\psi_s$ vs $V_g$ curves for  different thicknesses intersect at $\psi_s$=0. The corresponding value of $V_g$ is the flatband voltage $V_{fb}$, which, in principle, should be independent of thickness of the semiconductor. 

The total capacitance $C$, calculated using $Q_{sp}$, $\psi_s$ and \eqref{EqTotalC}, is plotted with respect to $V_g$ for different semiconductor thicknesses, as shown in Fig. \ref{Fig2CV_TCAD_analytical}(a). Excellent agreement between the analytical model (solid lines) and TCAD simulation results (symbols), in all the figures (Figs. \ref{Fign_E_Q_Vs} and \ref{Fig2CV_TCAD_analytical}) ensures the consistency, scalability and robustness of the proposed model. Moreover, the model captures the physics behind $C$-$V$ characteristics over the entire voltage range, covering regions 1-3. It is noteworthy to mention that the variation of $C_{min}$ and slope of $C$-$V$ characteristics in the moderate accumulation region  with respect to semiconductor thickness is quite different from traditional MIS capacitor consisting of doped semiconductor. 	
	
	\section{Experimental Results}
	We extend our study to validate the present model with experimental data. Organic semiconductor based MIS capacitors were fabricated on cleaned glass substrates by evaporating aluminum of thickness 50 nm as gate metal, followed by spin coating of cross linked Poly(4-vinylphenol) (PVP) with spin speed 6000 r$/$min. Subsequently, the samples were annealed at 200 $^oC$ for 20 minutes that resulted in a thickness of 290 nm (measured using an ellipsometer). Poly(3-hexylthiophene-2,5-diyl) (P3HT), a widely studied undoped organic semiconductor, was spun from a solution of concentration 15 mg$/$ml in 1,2-Dichlorobenzene. The spin speed was set to 1000, 3000 and 4000 r$/$min to achieve P3HT thickness of 97, 56 and 27 nm, respectively. The samples were then annealed at 120 $^oC$ for 20 minutes prior to thermal evaporation of gold of thickness 50 nm as the top metal contact (cathode in particular). All the fabrication processes were carried out inside the glove box with controlled oxygen ($<$ 1 ppm) and moisture ($<$ 1 ppm) levels. The fabricated devices were characterized inside a vacuum probe station using Agilent B1500A parameter analyzer at 1 kHz frequency.
	
	The measured $C$-$V$ characteristics of the fabricated organic MIS capacitors with different semiconductor thicknesses are shown in Fig. \ref{Fig2CV_TCAD_analytical}(b) (star symbol), exhibiting the similar trend qualitatively to $C$-$V$ characteristics described using TCAD simulation results and the analytical model in Fig. \ref{Fig2CV_TCAD_analytical}(a). The dielectric constants of PVP and P3HT were extracted using the measured thicknesses, $C_{max}$ and $C_{min}$, showing consistency with the reported values\cite{PVPRef}. Analyzing the proposed model and the measured data, $V_{fb}$ and $\phi_h$ were extracted to reproduce $C$-$V$ characteristics (lines) in Fig. \ref{Fig2CV_TCAD_analytical}(b). The same parameters were used for TCAD simulation results, as represented by the symbols in Fig. \ref{Fig2CV_TCAD_analytical}(b). Appreciable coherence between the experimental result (solid stars), TCAD simulation (symbols) and analytical model (lines) validates the applicability of the presented model in case of organic semiconductor (undoped) based MIS capacitors.	
	\section{Conclusion}
	In conclusion, the $C$-$V$ characteristics of undoped organic semiconductor based MIS capacitor is thoroughly investigated. Even though the nature of the characteristic seems very similar to that of doped semiconductor based MIS capacitor, there are distinct differences between the underlying device physics of doped and undoped semiconductor based MIS capacitors. A physics based model, incorporating the injection of mobile charge carriers through metal-semiconductor Schottky contact and their accumulation within the semiconductor under the influence of gate voltage, is developed to explain $C$-$V$ characteristics over entire voltage range. The model is consistently validated with TCAD simulation results. Variation of $C$-$V$ characteristics of organic MIS capacitor with the semiconductor thicknesses, which is unlikely in traditional MIS capacitor, is verified with the experimental data obtained from P3HT based MIS capacitor. 
	
	\section*{Acknowledgements}
	The authors would like to acknowledge Centre for NEMS and Nanophotonics (CNNP) for providing fabrication facility. Authors would like to thank Nidhin K. for fruitful discussion.
	
	%
	
	\bibliographystyle{ieeetr}

	%
	
	%
	%
	%
	
	
	

\end{document}